# Macroscopic entanglement between ferrimagnetic magnons and atoms via crossed optical cavity

Ke Di [1], Xi Wang [1], Huarong Xia [1], Yinxue Zhao [2], Anyu Cheng [1], Yu Liu [1], and Jiajia Du [1*]

[1]Chongqing University of Post and Telecommunications, Chongqing, 400065, China
[2]Wuhan Social Work Polytechnic, Wuhan, 430079, China

E-mail: *dujj@cqupt.edu.cn*

**Abstract**

We consider a two-dimensional opto-magnomechanical (OMM) system including two optical cavity modes, a magnon mode, a phonon mode, and a collection of two-level atoms. In this study, we demonstrate the methodology for generating stationary entanglement between two-level atoms and magnons, which are implemented using two optical cavities inside the setup. Additionally, we investigate the efficiency of transforming entanglement from atom-phonon entanglement to atom-magnon entanglement. The magnons are stimulated by both a bias magnetic field and a microwave magnetic field, and they interact with phonons through the mechanism of magnetostrictive interaction. This interaction generates magnomechanical displacement, which couples to an optical cavity via radiation pressure. We demonstrate that by carefully selecting the frequency detuning of an optical cavity, it is possible to achieve an increase in bipartite entanglements. Furthermore, this improvement is found to be resistant to changes in temperature. The entanglement between atoms and magnons plays a crucial role in the construction of hybrid quantum networks. Our modeling approach exhibits potential applications in the field of magneto-optical trap systems as well.

## 1. Introduction

The magnon mode, which refers to the quanta of collective spin excitation in yttrium iron garnet, has garnered significant interest owing to its notable characteristics such as high spin density, low damping rate, and tunability. Moreover, this remarkable characteristic is employed to establish a robust connection between the magnon mode in yttrium iron garnet (YIG) and the microwave cavity photon mode, enabling the realization of quantum information transmission inside a hybrid system [1,2]. Many experiments have confirmed the realization of cavity-magnon strong coupling within the YIG sphere [3,4]. Hybrid ferrimagnetic systems offer a novel framework for the manifestation of macroscopic quantum phenomena. These phenomena encompass a wide range of effects, including magnon dark modes and gradient memory [5,6], exceptional points [7,8], bistability of cavity-magnon polaritons [9], magnon cooling [10], nonreciprocity and unidirectional invisibility [11], Kerr effect in magnons [12,13], single-shot detection of individual magnons [14], and remote asymmetric quantum steering of magnons [15].

Cavity magnomechanics has similar properties to cavity optomechanics and hence enabling the exploration of several intriguing physical phenomena and their significant applications in the fields of quantum information science and quantum technology [16,17]. Cavity optomechanical system demonstrates strong coupling between the modes of the optical cavity and the vibrational modes via radiation pressure [18]. Based on the nature of cavity optomechanics and Cavity magnomechanics, the magnomechanical displacement couple to an optical cavity via radiation pressure [19-21]. The preparation of microwave-optics entanglement in a hybrid system was seen in accordance with a forementioned interaction [21]. Through an ensemble of two-level atoms coupled optical cavity, entangled states are obtained between atomic ensemble and magnon mode.

In this paper, we present a demonstration of the bipartite entanglement, denoted as $E_{ab}(E_{am})$ between a collective of two-level atoms and the phonon mode (also known as the magnon mode). Simultaneously, we calculated the entanglement transformation efficiency at different vertical optical cavity detuning. The approach we suggest is implemented in the context of quantum teleportation [22,23], quantum networks [24], quantum logical operations [25], quantum metrology [26], and fundamental tests of quantum mechanics [27].

The remainder of the paper is organized as follows. In Sec. 2, we introduce a hybrid opto-magnomechanical system, provide the Hamiltonian of the system, and derive the related quantum Langevin equations (QLEs). We next offer a comprehensive explanation of the methodology employed to solve the QLEs and compute the entanglement. In Sec.3, the primary findings about bipartite entanglement under varying vertical optical cavity detuning and entanglement transfer rate are presented. In conclusion, we arrive at our last remarks in Section 4.

## 2. The model

We consider a hybrid opto-magnomechanical system [21], which comprises of two optical cavity modes, a magnon mode, a phonon mode, and an ensemble of two-level atoms, as seen in Fig.1. The magnons are embodied by the collective motion of a large number of spins within a ferrimagnet [2]. The magnons couple to phonons via magnetostrictive interaction. The deformation of the geometric structure of the YIG crystal is a result of the magnetization variation caused by the excitation of magnons. This phenomenon is initiated by subjecting the crystal to a uniform bias magnetic field [1]. An ensemble of two-level atoms is positioned at the intersection of two optical cavities, and it is offresonantly coupled by a collective Tavis-Cummings-type interaction to the two optical fields [28,29]. The displacement of the YIG crystal couples to the horizontal optical cavity ($c_2$) via radiation pressure [30]. This is achieved by affixing a mirror pad, which is of micron size and highly reflective, onto the surface of the YIG micro bridge [20].

The Hamiltonian of the system under rotating-wave approximation in a frame rotating with the frequency of the drive field is given by

$$H_0 = \sum_{j=1,2} \hbar\omega_{cj} c_j^\dagger c_j + \hbar\omega_m m^\dagger m + \frac{\hbar\omega_b}{2}(q^2 + p^2)$$
$$+ (\hbar\omega_a S_Z)/2 + \hbar g_m m^\dagger m q + \hbar \sum_{j=1,2} g_a(c_j^\dagger a + a^\dagger c_j)$$
$$- \hbar g_c c^\dagger c q + H_{dri} \quad (1)$$

where m and $m^\dagger$ ($c_j$ and $c_j^\dagger$), [O, O$^\dagger$]=1, O = m ($c_j$), $\omega_c$, $\omega_m$, and $\omega_b$ are the resonance frequency of the optical cavity, magnon, and mechanical modes. The first (second) term describes the energy of the optical cavity mode (magnon modes). The magnon frequency $\omega_m = \gamma H$ is determined by the bias magnetic field $H$ where $\gamma/2\pi$ = 28 GHz/T is the gyromagnetic ratio. The third term denotes the energy of two mechanical vibration modes, and $q_j$ and $p_j$ ($[q_j, p_j]$ = i) are the dimensionless position and momentum of the vibration mode j, modeled as a mechanical oscillator. The atomic ensemble consists of $N_a$ two level atoms with natural frequency $\omega_a$. $S_{\pm,Z} = \sum_{j=1}^{N_a} \sigma_{\pm,Z}^{(i)}$ for i = 1, $N_a$ denotes collective spin operators, which satisfy the commutation relations $[S_+, S_-] = S_Z$ and $[S_Z, S_\pm] = -2S_\pm$. $\sigma_+, \sigma_-$ and $\sigma_Z$ are the spin -1/2 algebra of Pauli matrices. The coupling rate $g_m$ denotes the linear coupling between the microwave filed and the magnon mode, and $g_c$ represents denotes the bare optomechanical coupling strength. The atom cavity coupling constant is defined as g = $\mu\sqrt{\omega_c/(2\hbar\varepsilon_0 V_C)}$, $V_C$ is the horizontal optical cavity mode volume and µ is the dipole moment of the atomic transition. The last term is the driving Hamiltonian, $H_{dri} = i\Omega\hbar(m_1^\dagger e^{-i\omega_0 t} - m_1 e^{i\omega_0 t}) + \sum_{j=1,2} iE\hbar(c_j^\dagger e^{-i\omega_L t} - c_j e^{i\omega_L t})$, where E = $\mu\sqrt{2P_L \kappa_C/(\hbar\omega_L)}$ represents the coupling strength between the optical cavity and the laser drive field, with $P_L$ ($\omega_L$) being the power (frequency) of the laser, and $\kappa_C$ the cavity decay rate. The Rabi frequency $\Omega = \frac{\sqrt{5}}{4}\gamma\sqrt{N}B_0$[1] denotes the coupling strength of the drive magnetic field (with amplitude $B_0$ and frequency $\omega_0$) with the first magnon mode, where the total number of spins N=$\rho$V with $\rho$=4.22×10$^{27}$ m$^{-3}$ the spin density of the YIG and V the volume of the sphere[9].

We consider a simplified version of the Hamiltonian, which is effective in the low-excitation limit. When all the atoms are initially prepared in their ground state, so that $S_Z \simeq \langle S_Z \rangle \simeq -N_a$ and this situation will not significantly change due to the interaction with the optical cavity. This condition is met when the likelihood of excitation for a single atom is low. In this limit, the dynamics of the atomic polarization can be described in terms of bosonic operators.

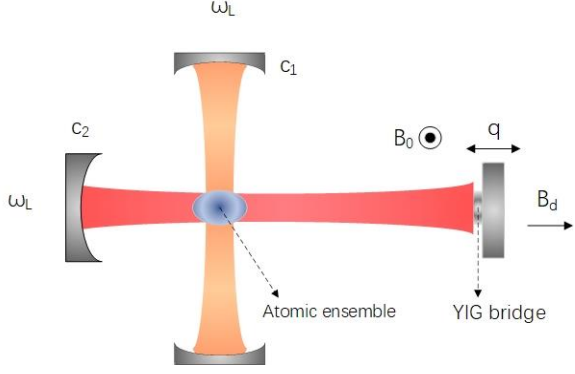

FIG. 1. Two optical cavity modes (c) driven by two lasers at frequency $\omega_L$ couples to an ensemble of two-level atoms (a), and a magnon mode (m) in a YIG crystal couple to a mechanical vibration mode (b) via magnetostriction. $B_0$ is the bias magnetic field, $B_d$ is the drive magnetic field.

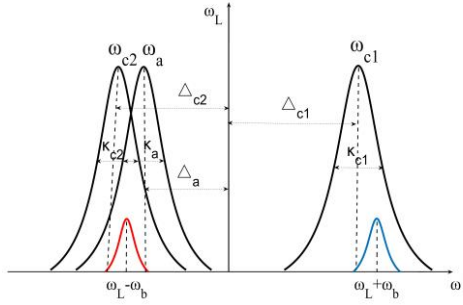

FIG.2. Frequencies and linewidths of the system. The ensemble of two-level atoms with frequency $\omega_a$ is driven by two optical fields at frequency $\omega_L$, and the mechanical motion of frequency $\omega_b$ scatters photons onto the two sidebands at $\omega_L \pm \omega_b$. If the optical cavity mode $c_2$ is resonant with the blue (anti-Stokes) sideband, and the optical cavity mode $c_1$ and atomic frequencies are resonant with the red (Stokes) sideband, a stationary atoms-magnon entangled state is generated.

In fact, $a = S_-/\sqrt{|\langle S_z\rangle|}$ is defined as the atomic annihilation operator, and it satisfies the usual bosonic commutation relation $[a, a^\dagger]=1$[31]. We obtain the fully simplified Hamiltonian, given by

$$H_0 = \sum_{j=1,2}\hbar\omega_{cj}c_j^\dagger c_j + \hbar\omega_m m^\dagger m + \frac{\hbar\omega_b}{2}(q^2+p^2)$$

$$+\hbar\omega_a a^\dagger a + \hbar g_m m^\dagger mq + \hbar\sum_{j=1,2}g_N(c_j^\dagger a + a^\dagger c_j)$$

$$-\hbar g_c c^\dagger cq + H_{dri} \quad (2)$$

where $g_N = g_a/\sqrt{N_a}$ denotes the effective atom-cavity coupling strength. The dissipative and fluctuation term are added to the Heisenberg equation of motion derived from the Hamiltonian. We establish a set of quantum Langevin equations of the system:

$$\dot{a} = -(i\Delta_a + \kappa_a)a - ig_{N1}c_1 - ig_{N2}c_2 + \sqrt{2\kappa_a}a^{in} \quad (3a)$$

$$\dot{c}_1 = -(i\Delta_{c1} + \kappa_{c1})c_1 + ig_c cq - ig_{N1}a + E + \sqrt{2\kappa_{c1}}c_1^{in} \quad (3b)$$

$$\dot{c}_2 = -(i\Delta_{c2} + \kappa_{c2})c_2 - ig_{N2}a + E + \sqrt{2\kappa_{c2}}c_2^{in} \quad (3c)$$

$$\dot{m} = -(i\Delta_m + \kappa_m)m - ig_m mq + \Omega + \sqrt{2\kappa_m}m^{in} \quad (3d)$$

$$\dot{q} = \omega_b p \quad (3e)$$

$$\dot{p} = -\omega_b q - \gamma_b p + g_c c^\dagger c - g_m m^\dagger m + \xi \quad (3f)$$

Where $\Delta_{a(c_j)} = \omega_{a(c_j)} - \omega_L$ and $\Delta_m = \omega_m - \omega_0$, $\kappa_j$, $\kappa_m$, and $\gamma_b$ are the dissipation rates of the cavity, magnon, mechanical modes, $\kappa_a$ is the decay rate of the two-level atoms excited level. Respectively, and $o^{in}$ (o=a, m, $c_j$) are zero-mean input noise operators, respectively, which are the correlation function is characterized as follows.

$$\langle o^{in}(t) o^{in\dagger}(t')\rangle = [N_o(\omega_o) + 1]\delta(t-t') \quad (4a)$$

$$\langle o^{in}(t') o^{in\dagger}(t)\rangle = N_o(\omega_o)\delta(t-t') \quad (4b)$$

The Langevin force operator $\xi$, which accounts for the Brownian motion of the mechanical mode, is autocorrelated as $\langle \xi(t)\xi(t') + \xi(t')\xi(t)\rangle/2 = \gamma_b[2N_b(\omega_b)+1]\delta(t-t')$[10], where we have made the Markov approximation, which is a good approximation for a mechanical oscillator of a large quality factor $Q_b = \omega_b/\gamma_b \gg 1$, and $N_k(\omega_k) = [\exp\left(\frac{\hbar\omega_k}{K_B T}\right)-1]^{-1}$, $K = a, c_j, m, b$, is the average number of thermal excitations for each mode at bath temperature T, respectively, with $K_B$ the Boltzmann constant.

he cavity mode driven by strong laser, YIG crystal driven by microwave magnetic field, which results in large steady-state amplitudes $|\langle c_j\rangle|, |\langle m\rangle| \gg 1$. This allows us to linearize the dynamics of the system around the steady-state values by writing any operator as $O = \langle O\rangle + \delta O$, (O=a, $c_j$, q, p, m), and neglecting small second-order fluctuation terms. Since we are particularly interested in the quantum correlation properties of the two magnon modes, we focus on the dynamics of the quantum fluctuations of the system. The linearized QLEs describing the fluctuations of the system quadratures $(\delta x_a, \delta y_a, \delta x_{c1}, \delta y_{c1}, \delta x_{c2}, \delta y_{c2}, \delta q, \delta p, \delta x_m, \delta y_m)$, with $\delta x_O = (\delta O + \delta O^\dagger)/\sqrt{2}$ and $\delta y_O = (\delta O^\dagger - \delta O)/\sqrt{2}$, can be written in the form of

$$\dot{u}(t) = Au(t) + n(t) \quad (5)$$

$$u(t) = [\delta x_a, \delta y_a, \delta x_{c1}, \delta y_{c1}, \delta x_{c2}, \delta y_{c2}, \delta q, \delta p, \delta x_m, \delta y_m]^T \quad (6)$$

$n(t) = [\sqrt{2\kappa_c}X^{in}(t), \sqrt{2\kappa_c}Y^{in}(t), \sqrt{2\kappa_1}x_1^{in}(t)\sqrt{2\kappa_1}y_1^{in}(t),$ $\sqrt{2\kappa_2}x_2^{in}(t), \sqrt{2\kappa_2}y_2^{in}(t), 0, \xi(t)]^T$ is the vector of input noises, and the drift matrix A is given by

$$A = \begin{pmatrix} -\kappa_a & \Delta_a & 0 & g_{n1} & 0 & g_{n2} & 0 & 0 & 0 & 0 \\ -\Delta_a & -\kappa_a & -g_{n1} & 0 & -g_{n2} & 0 & 0 & 0 & 0 & 0 \\ 0 & g_{n1} & -\kappa_{c1} & \Delta_{c1} & 0 & 0 & 0 & 0 & 0 & 0 \\ -g_{n1} & 0 & -\Delta_{c1} & -\kappa_{c1} & 0 & 0 & 0 & 0 & 0 & 0 \\ 0 & g_{n2} & 0 & 0 & -\kappa_{c2} & \Delta_{c2} & G_C & 0 & 0 & 0 \\ -g_{n2} & 0 & 0 & 0 & -\Delta_{c2} & -\kappa_{c2} & 0 & 0 & 0 & 0 \\ 0 & 0 & 0 & 0 & 0 & 0 & 0 & \omega_b & 0 & 0 \\ 0 & 0 & 0 & 0 & -G_C & 0 & -\omega_b & -\gamma_b & 0 & G_{mb} \\ 0 & 0 & 0 & 0 & 0 & 0 & -G_{mb} & 0 & -\kappa_m & \Delta_m \\ 0 & 0 & 0 & 0 & 0 & 0 & 0 & 0 & -\Delta_m & -\kappa_m \end{pmatrix}$$

(7)

where $\widetilde{\Delta_{c2}} \cong \Delta_{c2} - g_c\langle q\rangle$, and $\widetilde{\Delta_m} \cong \Delta_m + g_m\langle q\rangle$, represent the effective detuning of the optical cavity mode and magnon mode respectively, which include the frequency shifts due to the mechanical displacement $\langle q\rangle = (g_{c2}|\langle m\rangle|^2 + g_m|\langle m\rangle|^2)/\omega_b$, which includes the frequency shift due to the magnon-phonon interaction. $G_{mb} = i\sqrt{2}g_m\langle m\rangle$ is the effective magnomechanical coupling rate, $G_c = i\sqrt{2}g_{c2}\langle c_2\rangle$ is the effective optomechanical coupling rate, The average $\langle c_2\rangle$

$$\langle c_2\rangle = \frac{iE(g_{N1}g_{N1} - g_{N1}g_{N2} + (\Delta_a - i\kappa_a)(\Delta_{c1} - i\kappa_{c1}))}{g_{N2}g_{N2}(\Delta_{c1} - i\kappa_{c1}) + (g_{N1}g_{N1} + (\Delta_a - i\kappa_a)(\Delta_{c1} - i\kappa_{c1}))(\Delta_{c2} - i\kappa_{c2})} \quad (8)$$

The average $\langle m\rangle$ is given by

$$\langle m\rangle = \frac{\Omega}{(\kappa_m + i\Delta_{c2})} \quad (9)$$

Since the linearized dynamics and all noise are Gaussian, the system retains the Gaussian nature of any input state. The steady state of the quantum fluctuations of the system is a continuous variable five-mode Gaussian state, which is characterized by a 10×10 covariance matrix (CM) $V_{ij} = \langle u_i(t)u_j(t') + u_j(t)u_i(t')\rangle/2$ (i,j = 1,2,…,10). By solving the Lyapunov equation, it is straightforward to obtain the steady state CM [32,33].

$$AV + VA^T = -D \quad (10)$$

where $D = diag[\kappa_a, \kappa_a, \kappa_{c1}, \kappa_{c1}, \kappa_{c2}, \kappa_{c2}, 0, \gamma_b(2N_b + 1), \kappa_m(2N_m + 1), \kappa_m(2N_m + 1)]^T$ is the diffusion matrix, which is defined by $D_{ij}\delta(t - t') = \langle n_i(t)n_j(t') + n_j(t)n_i(t')\rangle/2$). We adopt the logarithmic negativity to quantify the magnon entanglement, The logarithmic negativity is defined as [34,35]

$$E_N \equiv max[0, -ln2\widetilde{v_-}] \quad (11)$$

where $\widetilde{v_-} = 2^{-1/2}\{\Sigma(V) - [\Sigma(V)^2 - 4detV_0]^{1/2}\}^{1/2}$, $V_0$ is the 4 × 4 CM associated with the two modes. $V_0 = [V_1, V_{12}; V_{12}, V_2]$, with $V_1, V_{12}$ and $V_2$ being the 2 × 2 blocks of $V_0$, and $\Sigma V \equiv detV_1 + detV_2 - 2detV_{12}$.

## 3. Analysis and discussion

Due to the highly thermally populated of mechanical modes at low temperatures, we prepare quantum states in the hybrid system by cooling mechanical modes to the ground state [36]. We drive the horizontal optical cavity ($c_2$) is operated by driving it with a red-detuned laser. This laser induces optomechanical anti-Stokes scattering, which effectively cools the mechanical modes in the sideband limit ($\omega_b \gg \kappa_{c2}$)[37]. When the optical cavity is driven by a strong laser, the weak coupling condition ($\omega_b \gg G_c$) is no longer valid. Therefore, we take the counter-rotating-wave (RW) terms $\propto c^\dagger b^\dagger + cb$ to obtain the parametric down-conversion (PDC) interaction, generate the optomechanical entanglement [33]. The initial cavity-phonon entanglement is partially redistributed to the atom-phonon and atom-magnon subsystems when the atomic frequency matches the Stokes sideband. Driving magnon modes with a red-detuned microwave field with detuning $\Delta_m = \omega_b$, the magnon frequency consistent with the anti-Stokes sideband, which realizes the magnon-phonon state-swap interaction [1].

We used the experimentally feasible parameter[38-41]: $\omega_m/2\pi = 10$ GHz, $\omega_b/2\pi = 40$ MHz, $\gamma_b/2\pi = 10^2$ Hz, $\kappa_a/2\pi = \kappa_m/2\pi = 1$ MHz, $\kappa_{c1}/2\pi = \kappa_{c2}/2\pi = 2$ MHz, $g_{N2}/2\pi = 8$ MHz, $G_m/2\pi = 2.5$ MHz, $G_c/2\pi = 8$ MHz, $\lambda_{L1} = \lambda_{L2} = 1064$ nm (optical wavelength), $\Delta_m = \omega_b$.

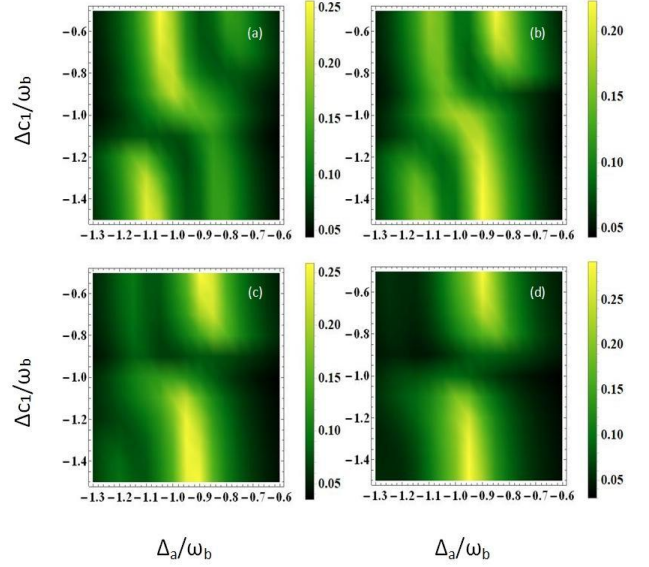

FIG.3. Density plot of bipartite entanglement between atoms and phonon ($E_{ab}$) versus dimensionless detunings $\Delta_a/\omega_b$ and $\Delta_{c1}/\omega_b$. (a) $\Delta_{c2} = -0.8\omega_b$, (b) $\Delta_{c2} = -0.9\omega_b$, (c) $\Delta_{c2} = -1.1\omega_b$, (b) $\Delta_{c2} = -1.2\omega_b$. We take $g_{N1}/2\pi = 4\ MHz$ and in all plots.

The strong optomechanical coupling is possible with a laser power $P_L = 4.4\ mW$. The magnomechanical coupling is realized by a microwave drive power $P_0 = 1.44\ mW$ for a 10μm³ YIG crystal with $g_m/2\pi = 20\ Hz$[6]. All values of the parameters are guaranteed by the negative eigenvalues (real part) of the drift matrix A.

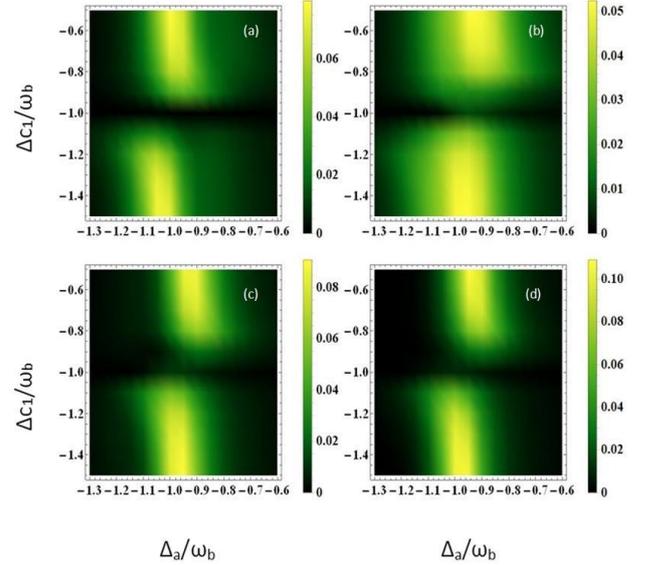

FIG.4. Density plot of bipartite entanglement between atoms and magnon ($E_{am}$) versus dimensionless detunings $\Delta_a/\omega_b$ and $\Delta_{c1}/\omega_b$. (a) $\Delta_{c2} = -0.8\omega_b$, (b) $\Delta_{c2} = -0.9\omega_b$, (c) $\Delta_{c2} = -1.1\omega_b$, (b) $\Delta_{c2} = -1.2\omega_b$. We take $g_{N1}/2\pi = 4\ MHz$ and in all plots.

The generation of entanglement in optomechanical systems is initiated, and subsequently spread to the atom-phonon system by the strength of the atom-cavity coupling. Furthermore, it is further distributed to the atom-magnon system through the strength of the magnomechanical nonlinear coupling. The modulation of entanglement between atoms and magnons is achieved by adjusting the detuning of the vertical optical cavity ($c_1$). We show the density plot of bipartite entanglement of atom-phonon versus $\Delta_a/\omega_b$ and $\Delta_{c1}/\omega_b$. Fig.3 indicates that the initial atom-phonon entanglement stems from the mechanical displacement and atom couple to the optical cavity. The anti-crossing around $\Delta_{c1} = \omega_b$ and $\Delta_a = -\omega_b$ in the figure is a signature of the strong coupling [30]. Figure 4 illustrates the entanglement $E_{am}$ between the atoms and the magnon mode, which is influenced by two optical cavities, as a function of dimensionless detunings $\Delta_a/\omega_b$ and $\Delta_{c1}/\omega_b$[42].

It is important to acknowledge that the detuning of the vertical optical cavity ($c_1$) significantly influences the entanglement in parameter space. The fluctuation of entanglement in the parameter range for varied detuning in the vertical optical cavity ($c_1$) is seen in Fig. 3 and Fig. 4. Transmission efficiency, denoted as the ratio $E_{am}/E_{ab}$, is also computed at different detuning of the vertical optical cavity ($c_1$)[43].

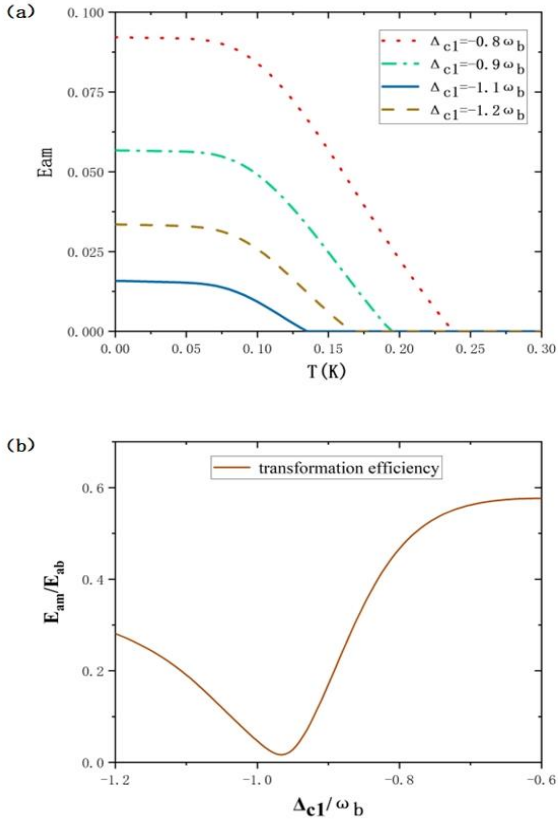

FIG. 5. (a) $E_{am}$ versus temperature T at different optical cavity ($c_1$) detunings. (b) Transformation efficiency versus optical cavity ($c_1$) detunings. We take $\Delta_{c1} = -0.8\omega_b$ (dotted curve), $\Delta_{c1} = -0.9\omega_b$ (dot dash curve), $\Delta_{c1} = -1.1\omega_b$ (solid curve), $\Delta_{c1} = -1.2\omega_b$ (dashed curve) in Fig. 5(a) and optimal detunings $\Delta_{c2} = -0.8\omega_b$, $\Delta_a = -0.95\omega_b$ in both plots.

In Fig. 5(a), the robustness of entanglement against temperature is altered by different optical cavity detuning. The atom-magnon entanglement is robust against thermal noises and is present at an environmental temperature up to T = 240mK. Strong entanglement exists at higher temperatures. As $\Delta_{c1}$ increases, Figure 5(b) illustrates that the transformation efficiency hits its minimum at approximately $\Delta_{c1} = -\omega_b$ and begins to climb. This is due to the fact that the robustness of entanglement against temperature decreases with $\Delta_{c1} \cong -\omega_b$ (Fig. 5(a), solid curve).

## 4. Conclusions

In this study, we provide a proposed methodology for the creation of entanglement between magnons and atoms inside an opto-magnomechanical (OMM) system. This is accomplished by employing two mutually perpendicular optical cavities to drive a collection of two-level atoms. It has been demonstrated that within the experimentally viable parameter ranges, the initial entanglement between photo-phonons is successfully transmitted to the subsystems of atomic-phonons and atomic-magnons. Additionally, we investigate the effectiveness of entanglement transmission between atom-phonon entanglement and atom-magnon entanglement. The establishment of atom-magnon entanglement arises from the synergistic utilization of opto- and magnomechanical cooling techniques, as well as optomechanical parametric down-conversion (PDC) interactions. The utilization of our modeling techniques has significant potential for application in magneto-optical trap systems [44].

This work was supported by National Natural Science Foundation of China (Grant Nos. 11704053, 52175531); the National Key Research and Development Program of China (Grant No. 2021YFC2203601).

AUTHOR DECLARATIONS
Conflict of Interest
The authors have no conflicts to disclose.

DATA AVAILABILITY
The data that support the findings of this study are available from the corresponding author upon reasonable request.